\documentclass[11pt,onecolumn,twoside]{IEEEtran}

\usepackage{listings}
\usepackage{caption}
\usepackage{subcaption, mwe}
\usepackage{makecell}

\usepackage{amssymb,amsmath, amsthm}
\usepackage{bm}
\usepackage{float}
\usepackage{graphicx}
\usepackage[htt]{hyphenat}
\usepackage{algorithm}
\usepackage{algpseudocode}
\usepackage{xcolor}
\usepackage{siunitx}
\usepackage{colortbl}
\usepackage{multicol}
\usepackage{enumitem}
\usepackage{fullpage}
\usepackage{cite}
\usepackage{mathtools}
\usepackage{booktabs}

\newtheorem{theorem}{Theorem}

\algrenewcommand\algorithmicrequire{\textbf{Input:}}
\algrenewcommand\algorithmicensure{\textbf{Output:}}

\newcommand{\rightedge}{ \negmedspace \rightarrow \negthinspace}

\newcolumntype{P}[1]{>{\centering\arraybackslash}p{#1}}

\newcommand{\dt}{digital twin}
\newcommand{\dts}{digital twins}

\definecolor{s_color}{HTML}{432818}
\definecolor{o_color}{HTML}{AB6B47}
\definecolor{d_color}{HTML}{375f8f}
\definecolor{q_color}{HTML}{3c3672}
\definecolor{r_color}{HTML}{3a7a68}
\definecolor{u_color}{HTML}{6b3d7c}
\definecolor{u_in_color}{HTML}{8f4d6c}
\definecolor{o_in_color}{HTML}{AB6B47}
\definecolor{n_color}{HTML}{f25c5c}

\definecolor{h_color}{HTML}{051F37}
\definecolor{varsigma_color}{HTML}{AA4B37}

\definecolor{varsigma_in_color}{HTML}{aa7337}
\definecolor{z_color}{HTML}{267474}
\definecolor{epsilon_color}{HTML}{8373ae}
\definecolor{c_color}{HTML}{BC83A9}
\definecolor{c_in_color}{HTML}{892d6b}

\lstdefinelanguage{dict}{
    morekeywords={},
    sensitive=false, 
    morecomment=[l]{//}, 
    morecomment=[s]{/*}{*/}, 
    morestring=[b]"
}

\lstdefinelanguage{tla}{
    morekeywords={let, if, then, else,
        in, except, unchanged,
        null, append  },
    sensitive=false, 
    morecomment=[l]{//}, 
    morecomment=[s]{/*}{*/}, 
    morestring=[b]" 
}

\lstdefinestyle{mystyle}{
    basicstyle=\ttfamily\footnotesize,
    breaklines=true,                 
    captionpos=b,                    
    keepspaces=true,                 
    numbersep=5pt,                  
    showspaces=false,                
    showstringspaces=false,
    showtabs=false,                  
    tabsize=2,
    framesep=5pt,
    xleftmargin=5pt,
    xrightmargin=5pt,
}

\begin{document}

\title{Formal Verification of Digital Twins with TLA and Information Leakage Control}

\author{Luwen Huang, Lav R.\ Varshney, and Karen E.\ Willcox%
\thanks{L.~Huang is with the Department of Computer Science, University of Texas at Austin (e-mail: luwen@cs.utexas.edu).}
\thanks{L.~R.\ Varshney is with the Department of Electrical and Computer Engineering and Coordinated Science Laboratory, University of Illinois Urbana-Champaign, USA (e-mail: varshney@illinois.edu).}
\thanks{K.~E.~Willcox is with the Oden Institute for Computational Engineering \& Sciences, University of Texas at Austin (e-mail: kwillcox@oden.utexas.edu).}
}%

\maketitle

\begin{abstract}
Verifying the correctness of a digital twin provides a formal guarantee that the \dt{} operates as intended. Digital twin verification is challenging due to the presence of uncertainties in the virtual representation, the physical environment, and the bidirectional flow of information between physical and virtual. A further challenge is that a \dt{} of a complex system is composed of distributed components. This paper presents a methodology to specify and verify \dt{} behavior, translating uncertain processes into a formally verifiable finite state machine. We use the Temporal Logic of Actions (TLA) to create a specification, an implementation abstraction that defines the properties required for correct system behavior. Our approach includes a novel weakening of formal security properties, allowing controlled information leakage while preserving theoretical guarantees. We demonstrate this approach on a \dt{} of an unmanned aerial vehicle, verifying synchronization of physical-to-virtual and virtual-to-digital data flows to detect unintended misalignments.
\end{abstract}

\section{Introduction}
\thispagestyle{empty}

This paper describes a formal methodology to design and model a \dt{} and prove its correctness properties using the Temporal Logic of Actions (TLA). We employ the National Academies' definition: ``A digital twin is a set of virtual information constructs that mimics the structure, context, and behavior of a natural, engineered, or social system (or system-of-systems), is dynamically updated with data from its physical twin, has a predictive capability, and informs decisions that realize value. The bidirectional interaction between the virtual and the physical is central to the digital twin''  \cite{nasem}. 

\begin{figure}
    \centering
    \includegraphics[width=4in]{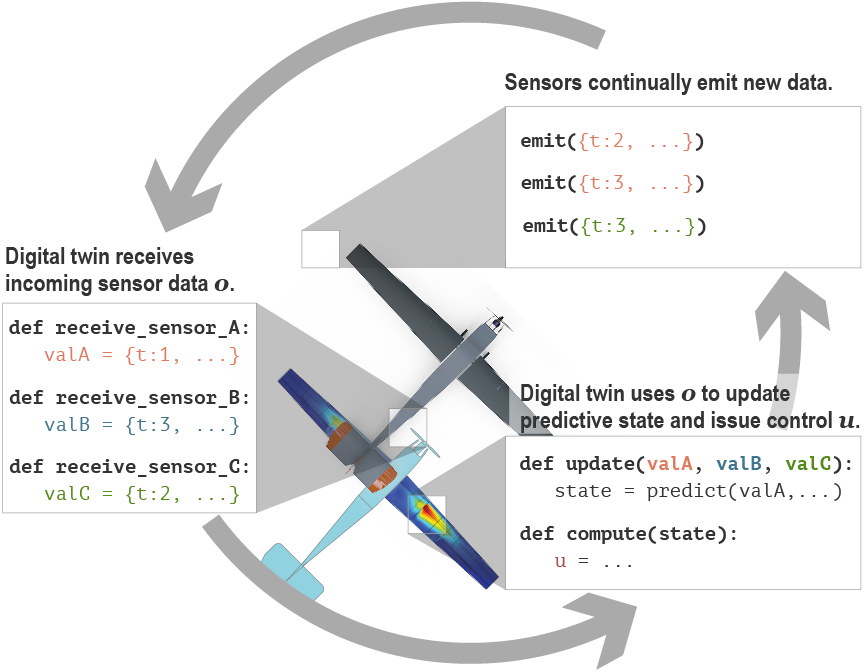}
    \caption{Digital twin consisting of a physical system (unmanned aerial vehicle), a virtual representation (structural health models), and bidirectional connections among components.}
    \label{fig:intro__uav-dt}
\end{figure}

An example \dt{} is illustrated in Fig.~\ref{fig:intro__uav-dt}. In this scenario, an unmanned aerial vehicle (UAV) flies a mission while transmitting sensor data to a \dt{}. The \dt{}, designed to mirror the UAV's structure, context and behavior, processes the incoming sensor data, $\bm{o}$, maintains a predictive model of the UAV's structural health, $\bm{s}$, and generates control execution commands, $\bm{u}$. Even when individual components, like the command-generation function, operate correctly, the system can still fail due to orchestration issues. For instance, consider the sensor observations {\color[HTML]{e07a5f} \texttt{valA}}, {\color[HTML]{40758E} \texttt{valB}}, and {\color[HTML]{568A19} \texttt{valC}}, arriving concurrently but with different timestamps {\color[HTML]{E07A5F} \texttt{t=1}}, {\color[HTML]{40758E} \texttt{t=3}}, and {\color[HTML]{568A19} \texttt{t=2} }, respectively. These readings, emitted at different intervals and with transmission delays, create orchestration challenges. To ensure the correct state, the \texttt{update} function must incorporate consistent sensor inputs, and the UAV must process incoming commands reliably to follow its intended path. These examples highlight that verifying individual component correctness is not enough; ensuring the \dt{}’s overall orchestration is equally important.

Various formal and technological approaches address aspects of correctness in cyber-physical systems. For instance, verifying control and timing is well-researched (see Sec.~\ref{sec:related-work}), but control verification alone does not ensure system-level orchestration. Technological solutions, such as RabbitMQ for asynchronous data handling, address only specific areas of \dt{} functionality. Moreover, simply adding technological components does not offer formal guarantees, often a crucial need in safety-critical environments where \dts{} may be deployed. This paper introduces a novel methodology for formally reasoning about \dts{} at the level of system orchestration. We introduce the following innovations:

\begin{enumerate}[leftmargin=*]
    \item \textbf{Formal system specification}: A new method to construct formal, high-level specifications of \dts{} using TLA. Our approach derives a finite state machine model from the \dt{} probabilistic graphical model (PGM) \cite{kapetyn2021}, giving a mathematically rigorous way to specify \dts{} in general.
    \item \textbf{Model augmentation}: A novel augmentation of the \dt{} PGM framework to model distributed communication and the corresponding state machine translation. 
    \item \textbf{Abstraction methodology}: A set of principled guidelines for abstracting the physical and computational complexities of \dts{} into state transition actions.

    \item \textbf{Weakening of formal properties}: A novel approach to relax formal security properties, such as non-interference, by bounding the utility of revealed information within \dt{} bidirectional flows, thereby limiting impact on system identification rather than relying on generic information-theoretic bounds.
\end{enumerate}

The remainder of this paper is organized as follows. Sec.~\ref{sec:related-work} places our approach in the existing literature. Sec.~\ref{sec:dt-state-machine} details the state machine derivation. Sec.~\ref{sec:application} demonstrates a practical application by constructing and verifying a UAV \dt{}, with relaxed security properties that provide formal bounds on information leakage between the physical and digital components. Finally, Sec.~\ref{sec:evaluation} presents the results of our verification efforts on the UAV \dt{}.

\section{Related Work} 
\label{sec:related-work}

Our research contributes to the field of cyber-physical systems, with particular focus on the expanding concept of \dts. As \dt{} technology continues to evolve rapidly, it is important to delineate how our approach both aligns with and diverges from existing work. 

\subsubsection*{Digital Twin frameworks}

Various works have suggested \dt{} design approaches that range from informal, flow chart-based design \cite{segovia}, \cite{erkoyuncu} to technology-specific solutions \cite{aboElHassan, temperekidis}. Unlike these approaches, our methodology is technology-neutral. Our approach offers a generalizable abstraction for \dt{} design that is grounded in mathematically rigorous formal verification principles.

\subsubsection*{Verification of cyber-physical systems} 

Verification of cyber-physical systems is a dynamic and expansive area of research \cite{mitra}, with much work in safe autonomy and control \cite{alur, vasile, kress-gazit, lee2017Book, platzer, sirjani, topcu2022,topcu2023, wright}. There is also considerable discourse on the challenges of distributed cyber-physical systems \cite{lee2008, lee2015, topcu2021}. For example, \cite{bateni} extends the Lingua Franca coordination language to handle network failures, \cite{mubeen, shrivastava} discuss the need for timing considerations in distributed environments, and \cite{davis, ghosh, mehrabian} offer methods to achieve deterministic timing in control executions. In contrast, our paper provides a methodology for orchestrating \dts{} at the system level. Aligning with the NASEM definition of a \dt, our approach emphasizes the critical importance of bidirectional interactions and orchestration, offering a broader, systemic perspective that diverges from the control-centric emphasis found in much existing literature.

\subsubsection*{Distributed systems}

The application of TLA in distributed computing systems is well-documented \cite{merz,howard, konnov, regnier}, with notable applications including its use at Amazon Web Services for managing distributed resources \cite{newcombe}. While TLA has proven effective in addressing the complexities of distributed computing, the specification of \dts{} presents unique challenges that extend beyond traditional distributed systems: First, \dts{} require the consideration of diverse hardware components, which goes beyond the typical software and network considerations found in distributed systems \cite{lee2008}. Second, \dts{} often incorporate predictive models that provide probabilistic outputs and may adapt dynamically based on real-time data. Third, \dts{} necessitate continual, real-time bidirectional exchanges to maintain synchronization between the physical and digital entities. Our research applies TLA to address these aspects, offering a formal, verifiable system perspective for \dts{}. To our knowledge, this represents a novel application of TLA in the context of \dts{}.  Moreover, we provide a systems-theoretic approach to give formal statistical guarantees on information leakage during communication between the physical system and its \dt{}.

\section{Digital Twin as a State Machine} 
\label{sec:dt-state-machine}

Our first result is formalizing a \dt{} as a state machine, rigorously derived from the \dt{} Probabilistic Graphical Model (PGM) framework proposed in \cite{kapetyn2021} and since adopted to describe \dts{}  in a variety of applications \cite{chaudhuri2023predictive, torzoni}. Appendix~\ref{appendix:background} provides a background of the \dt{} PGM framework. Our formalization uses TLA to describe the \dt{} as a finite state machine (FSM). Indeed, Markov models as in PGMs are a stochastic version of FSMs. For background on TLA, see \cite{lamport1977, lamport1994, lamport2002}. Throughout this section, we use examples from our application instance of a UAV \dt. 
However, we emphasize and show that our methodology is broadly applicable. 

\subsection{State Machine Derivation}

Here we detail our novel derivation of a state machine representation from the \dt{} PGM. We specify the \dt{} as a state machine that transitions from one state to the next, governed by transition logic:
\begin{equation}
        \textrm{Digital Twin} \coloneq \mathcal{I} \land \mathcal{N} \land \mathcal{F}
    \label{eqn:dt-as-state-machine__initial-next}
\end{equation}
Here, \eqref{eqn:dt-as-state-machine__initial-next} states the state machine of the \dt{} is defined by a conjunction (\texttt{AND}) of an initial state predicate $\mathcal{I}$, a next state predicate $\mathcal{N}$, and a set of fairness conditions $\mathcal{F}$. The initial state predicate specifies the valid starting conditions, the next state predicate outlines the permissible transitions that variables can undergo, and the fairness conditions provide assumptions about how transitions are executed. Specifically, the next state predicate:
\begin{equation}
    \begin{split}
        \mathcal{N} \coloneq \omega_1 \lor \cdots \lor \omega_N \lor \mathcal{T}
    \end{split}
    \label{eqn:dt-as-state-machine__next}
\end{equation}
employs the logical disjunction (\texttt{OR}) to indicate that at any given step in the state machine, one out of $N$ possible processes $\omega_n$ can occur, or the system can reach a termination condition  $\mathcal{T}$. By allowing only one process to execute at a time, we model concurrency by considering the possible orderings, or interleavings, of process execution, abstracting away timing specifics.

To model \dt{} orchestration, we identify the specific operations or ``processes'' through which variables within the system alter their states. These processes are dictated by the relationships encoded within the DT PGM. Each variable $v_i$ in the model transitions based on the states of other variables that directly influence it --- the variable node's parents in the graphical model. Formally, the set of processes $\Omega$ is defined as:
\begin{equation}
    \label{eqn:model__initial-process-set}
    \begin{split}
        \Omega = \{\textrm{Define} \; \omega \coloneq W(v_i) \rightarrow v_i \mid v_i \in V \land W(v_i) \neq \varnothing \} \\
    \end{split}
\end{equation}
Here, $W(v_i)$ is the set comprising the parents of $v_i$, and the transition function $\rightarrow$ denotes the computation that updates $v_i$ based on these influences. This definition preserves the system dependencies by defining that each variable's change is a direct result of its process' inputs.

\begin{figure}
    \includegraphics[width=4in]{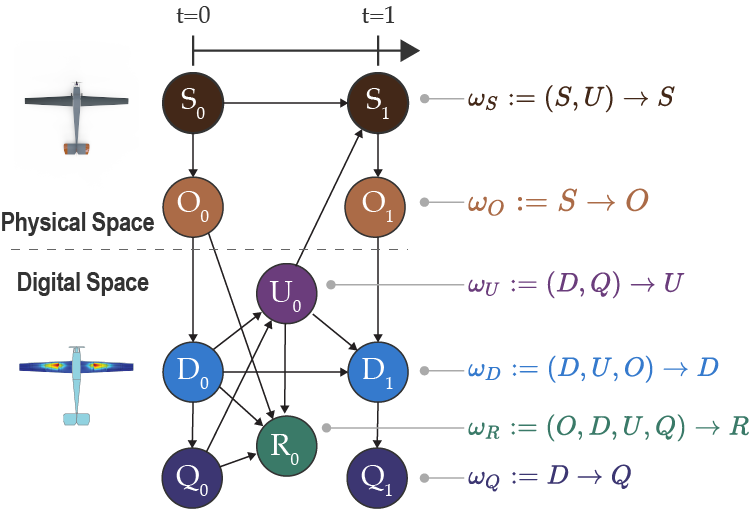}
    \centering
    \caption{Derivation of state machine processes from PGM representation. Nodes represent variables and edges between nodes represent the dependence of the destination node on the parent node. The subscript $X_t$ denote the variable $X$'s state at time $t$.}
    \label{fig:model__pgm-processes}       
\end{figure}

Fig.~\ref{fig:model__pgm-processes} shows the PGM describing our example UAV \dt{} with six variables: (1) physical state $S$ which represents the structural health of the UAV; (2) Observational data $O$ representing sensor data; (3) Digital state $D$, which represents the \dt{}'s estimate of the UAV's structural health; (4) control $U$ representing the computed control; (5) Quantity of interest $Q$, which represents quantities of interest computed by the \dt{}; and (6) Reward $R$, representing metrics for success as dependent on $O$, $D$, $U$ and $Q$.
Applying \eqref{eqn:model__initial-process-set} to the PGM yields the set of processes $\Omega = \textcolor{s_color}{\omega_S} \lor \textcolor{o_color}{\omega_O} \lor \textcolor{u_color}{\omega_U} \lor \textcolor{d_color}{\omega_D} \lor \textcolor{r_color}{\omega_R} \lor \textcolor{q_color}{\omega_Q}$ where $\textcolor{s_color}{\omega_S \coloneq (S,U) \rightarrow S}$, $\textcolor{o_color}{\omega_O \coloneq S \rightarrow O}$, $\textcolor{u_color}{\omega_U \coloneq (D,Q) \rightarrow U}$, $\textcolor{d_color}{\omega_D \coloneq (D, U, O) \rightarrow D}$, $\textcolor{r_color}{\omega_R \coloneq (O, D, U, Q) \rightarrow R}$ and $\textcolor{q_color}{\omega_Q \coloneq D \rightarrow Q}$. Fig.~\ref{fig:model__pgm-processes} illustrates the mapping of PGM encodings to state machine processes.

\subsection{Modeling Distributed Communication}
The second major contribution of our work is the novel augmentation of the \dt{} PGM to account for the challenges of distributed comopnents, and the corresponding translation into the state machine representation.

\begin{figure}[t]
    \centering
    \includegraphics[width=2.5in]{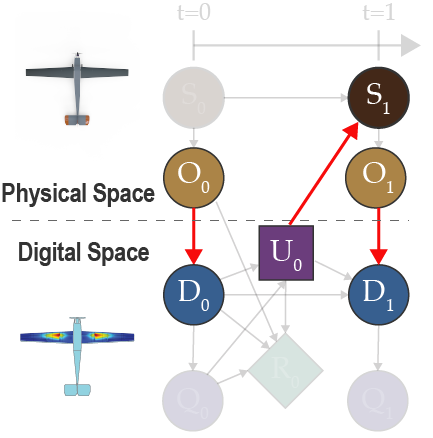}
    \caption{PGM with distributed communication required for two processes: (1) $O \rightedge D$ and (2) $U \rightedge S$.}
    \label{fig:model__pgm--distributed}
\end{figure}

The graphical model in Fig.~\ref{fig:model__pgm-processes} assumes that variable values are read deterministically. This is often not the case in \dts{} where components are distributed and rely on message passing to communicate with each other. With distributed components, there is additional \textit{uncertainty} in the input values that are actually used by a process, stemming from issues such as network reliability and traffic. For instance, as illustrated in Fig.~\ref{fig:model__pgm--distributed}, the process $O \rightedge D$ requires the value of $O$, which is transmitted via distributed messaging --- in this case, a wireless network channel.  The perturbation of the distributed messaging in a PGM might even be adversarial and so a worst-case analysis may be needed \cite{IhlerFW2005}.

\begin{figure}
    \centering
    \includegraphics[width=2.4in]{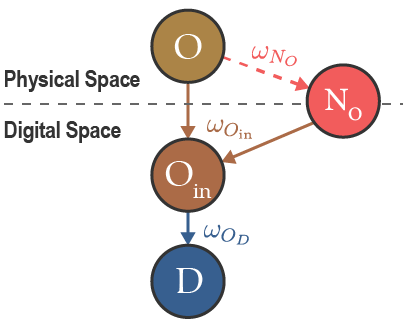}
    \caption{Augmentation for $O$, which is communicated over a distributed channel to $D$.}
    \label{fig:model__pgm--augmentation}
\end{figure}

Our novel augmentation of the PGM constructs a new variable $N$ to represent the uncertainty of the messaging channel and a new variable $X_\textrm{in}$ for every variable $X$ whose value is communicated over the messaging channel. These noise and channel output variables are just like in information-theoretic models of communication \cite{Shannon1948}, but considering semantics of logic \cite{BarHillelC1953}. First, identify the set of variables $\mathcal{X}$ whose value is communicated over distributed messaging, i.e. $\mathcal{X} = \{ X_i \mid X_i \in W(v_i) \land v_i \in V \land X_i \rightedge v_i \; \textrm{is distributed}  \}$. For every $X_i \in \mathcal{X}$, we create an intermediary variable $X_{i,\textrm{in}}$ and a network variable $N_{X_i}$ to represent the value of $X_i$ actually received. We reconfigure the incident edges of $X_i$ such that new edges point from $X_i$ to $X_{i_\textrm{in}}$, $N_{X_i}$ to $X_{i_\textrm{in}}$, $O$ to $N_{x_i}$, and $X_{i_\textrm{in}}$ to $V_i$. Algorithm~\ref{alg:pgm-transformation} details the augmentation algorithm.

For example, Fig.~\ref{fig:model__pgm--augmentation} shows a subgraph of the resulting augmentation applied to variable $O$. The augmentation introduces three new processes: (1) {\color[HTML]{F25C5C} $\omega_{N_O}$}, which represents the optional dependence of the messaging channel on the value of $O$. For instance, some messaging channels may be susceptible to large data payloads and may degrade as traffic increases. (2) { \color[HTML]{AB6B47} $\omega_{O_\textrm{in}}$ }, which represents the dependency of $O_\textrm{in}$ on both the message that was sent and the state of the network. (3) { \color[HTML]{375F8F} $\omega_{O_\textrm{in}}$ }, which replaces the original process $O \rightedge D$ to model the fact that the process input is the received variable $O_\textrm{in}$, instead of sent variable $O$. Our state machine formalization further elaborates on fairness, termination, and complexity abstraction, detailed in  Appendix~\ref{appendix:dt-state-machine}. 

\begin{algorithm}
\caption{Augment PGM for communication uncertainty}
\label{alg:pgm-transformation}
\begin{algorithmic}
\Require PGM $\mathcal{G}$
\Ensure Augmented PGM $\mathcal{G}'$
\State Set $V \gets$ set of nodes in $\mathcal{G}$
\State Set $E \gets$ set of edges in $\mathcal{G}$
\For{v in V} 
    \State Define $v \rightarrow w$ to be the outgoing edge from $v$ to node $w$
    \If{$v \rightarrow w$ \textrm{is distributed}}
        \State Create $v_\textrm{in}$ as new node, $V \gets V \cup v_\textrm{in}$
        \State Create $n_v$ as new node, $V \gets V \cup n_v$
        \State Remove edge $v \rightarrow w$ from $E$
        \State Create new edge $v \rightarrow v_\textrm{in}$, $E = E \cup v \rightarrow v_\textrm{in}$
        \State Set new edge $n_v \rightarrow v_\textrm{in}$, $E \cup n_v \rightarrow v_\textrm{in}$
        \State Set new edge $v_\textrm{in} \rightarrow w$, $E \cup v_\textrm{in} \rightarrow w$
    \EndIf
\EndFor 
\State \textbf{return} V, E
\end{algorithmic}
\end{algorithm}

\section{Specification of UAV Digital Twin}
\label{sec:application}

This section applies our proposed methodology to the design, specification, and verification of a UAV \dt. 

\subsection{The UAV and its Digital Twin}
\begin{figure}[t]
    \begin{subfigure}{.9\linewidth}
        \centering
        \includegraphics[width=\linewidth]{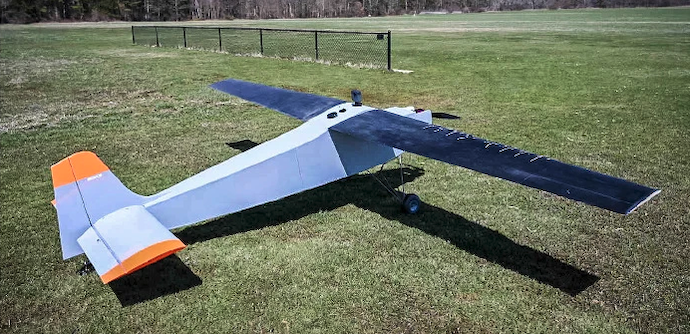}
        \caption{Testbed UAV}
    \end{subfigure}
    \vspace{.5cm}
    \begin{subfigure}{.9\linewidth}
        \centering
        \includegraphics[width=\linewidth]{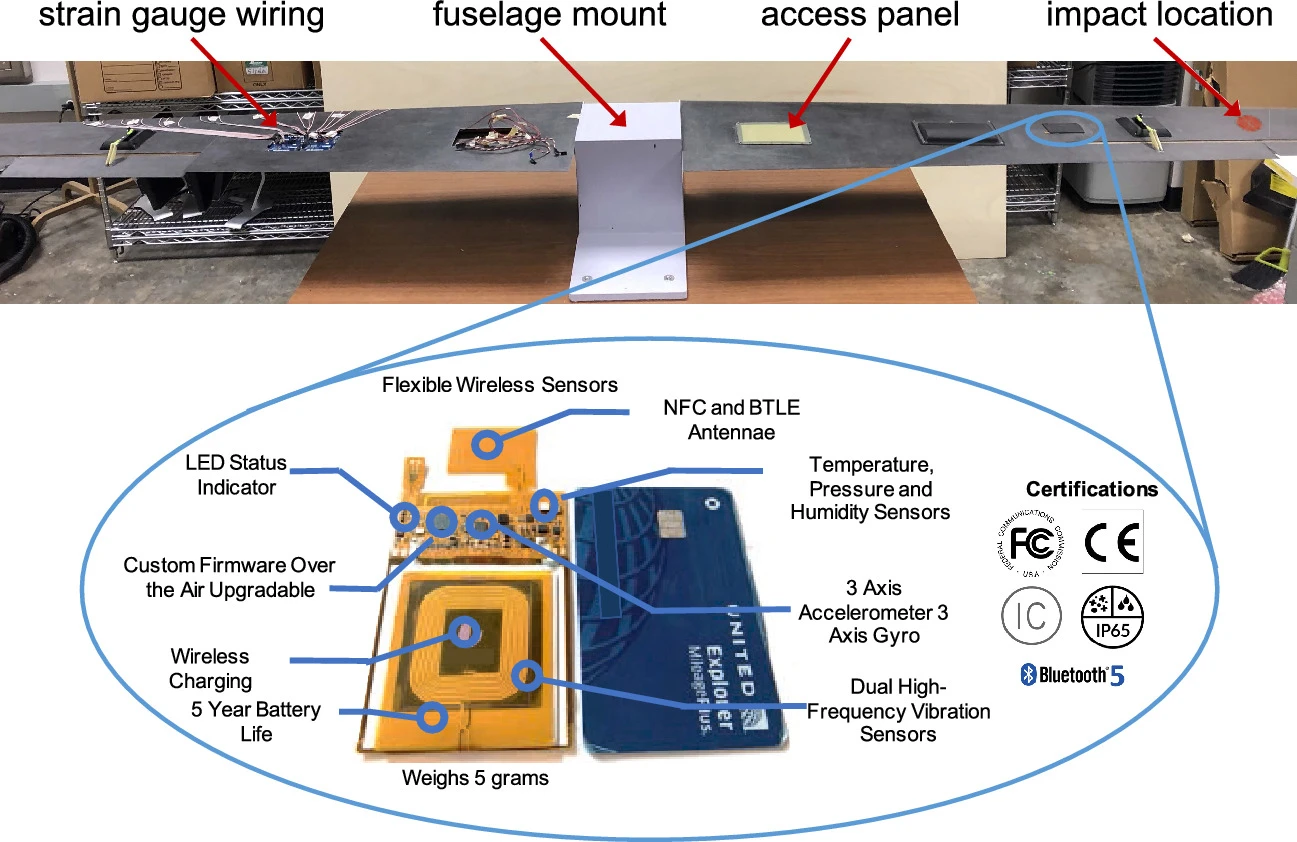}

        \caption{sensors on UAV wing}
        \label{fig:application__uav-sensors}
    \end{subfigure}
    \caption{Reproduced with permission from \cite{salinger}:  testbed UAV (top) equipped with individually-transmitting Bluetooth sensors (bottom)}
    \label{fig:application__uav}
\end{figure}

The physical counterpart of this \dt{} is a custom-built, fixed-wing UAV equipped with advanced wireless sensors and power hardware, with construction described in \cite{salinger} and shown in Fig.~\ref{fig:application__uav}. The sensors, attached to the UAV's wings as in Fig.~\ref{fig:application__uav-sensors}, measure observational data such as temperature and strain in real-time during flight. 
The UAV also features an onboard computer to process incoming control commands from the \dt, where commands are executed as maneuvers.
Given the potential unreliability of the 
communication channel, a primary design challenge is ensuring that delayed control messages are processed accurately to maintain the UAV's operational integrity.

A \dt{} of this UAV would continually process incoming observational data to generate and transmit control commands tailored for the UAV. The \dt{} would also maintain a dynamic predictive model of the UAV's structural health, ensuring synchronization with the UAV's actual physical state. This synchronization is achieved through real-time computations that integrate new observational data into the ongoing assessment of the UAV's condition. A key design challenge of the \dt{} is its ability to accurately reflect the UAV's physical state despite potential latency issues and concurrent, incoming data streams.

Our implementation builds upon the \dt{} in \cite{kapetyn2021}, implemented as a collection of Robot Operating System (ROS2) Python modules. However, the original implementation primarily served as a proof-of-concept for the PGM framework and did not address several real-world challenges such as handling concurrent incoming observational messages and ensuring reliability over unstable communication channels. Our objective is to construct a design to manage these complexities and achieve reliable orchestration under realistic operational conditions.

\subsection{The UAV State Machine}

\begin{figure}[t]
    \centering
    \includegraphics[width=4in]{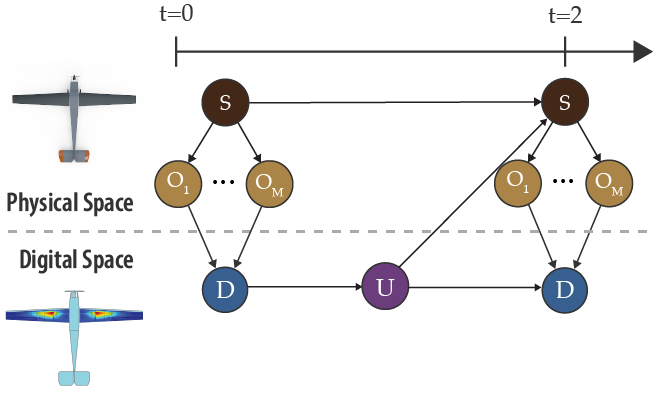}
    \caption{PGM for UAV Digital Twin}
    \label{fig:application__pgm}
\end{figure}

\begin{figure}[h]
    \centering
    \includegraphics[width=4.3in]{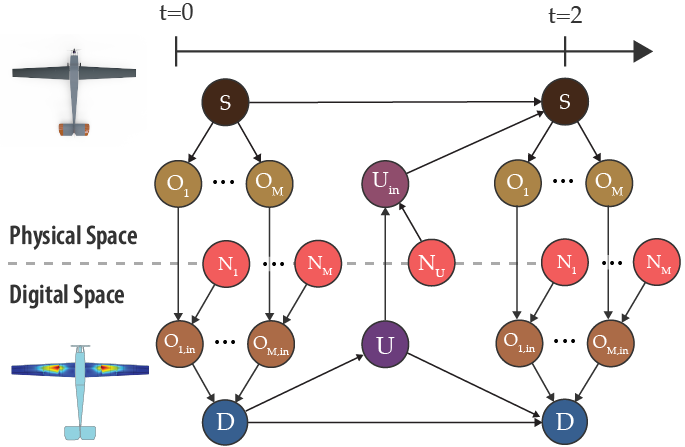}
    \caption{Augmented PGM modeling distributed communication }
    \label{fig:application__pgm-augmented}
\end{figure}

We apply our augmentation methodology from Algorithm~\ref{alg:pgm-transformation} to construct an augmented PGM that accounts for the distributed messaging channels in the system. Because each sensor transmits independently, we specify each sensor's connection as separate variables, $N_1 \ldots N_m$. We also define a separate variable for the transmission of control commands, $N_u$. These definitions let us reason about bidirectional flows individually.
In addition to new variables for each data transmission path, the augmentation also introduces new nodes for received observational data $O_{1,\textrm{in}} \ldots O_{m,\textrm{in}}$ and received control command $U_\textrm{in}$. The augmented PGM is depicted in Fig.~\ref{fig:application__pgm-augmented}. 

\subsection{System Abstraction}

This subsection applies our abstraction methodology to model concrete state transitions within the UAV \dt{}'s state machine. In abstracting complex system dynamics into simpler, formal state transitions, our goal is to balance fidelity with tractability: while a more granular formalization more accurately reflects real-world dynamics, it becomes less scalable in terms of formalization effort and verification time. We organize this section by systematically addressing each variable involved in the UAV system. For each variable, we first describe its real-world characteristics and then its corresponding abstraction. Following this, we delineate how each variable evolves in real life and how we formulate its state transition within the state machine.  \\

\noindent \textbf{Physical state ($S$)} The physical state $S$ of the UAV represents its structural health, which is influenced by the stresses of executed maneuvers. It is not possible for the \dt{} to know the ground truth of $S$ at runtime; instead, it must be inferred through sensor data. The UAV's structural integrity is subject to degradation, quantified by $\delta$ damage, which occurs with a non-zero probability dependent on the executed control. This probabilistic damage is governed by the dynamics shown in \eqref{eqn:application__physical-state__reality} in Table~\ref{tab:application__physical-state}.

In our abstraction, we model the structural health $S$ as a discrete variable ranging from $0$ (total structural failure) to $100$ (perfect health). Our model simplifies probabilistic damage to a nondeterministic state transition where the structural state either remains unchanged or is reduced by $\delta=1$ damage, shown in \eqref{eqn:application__physical-state__abstraction} in Table~\ref{tab:application__physical-state}. Finally, because damage occurs concurrently with control execution, both actions are modeled as a single atomic operation in \eqref{eqn:application__physical-state__abstraction} in Table \ref{tab:application__physical-state}, where the value of the next executed control $u_e$ is assigned the control command $u_\textrm{in}$.

\begin{table}[h!]
  \caption{Transition $(S, U) \rightedge S$: Evolution of physical state}
  \label{tab:application__physical-state}
\centering
    \begin{tabular}{P{3.8cm}|P{3.8cm}}
        \textbf{Real-world process} & \textbf{Abstraction} \\
        \midrule
            \begin{minipage}{3.8cm}
                \begin{equation}
                    \phi = \begin{cases}
                    0.05 & \textrm{if} \; u=3 \\
                    0.01 & \textrm{if} \; u=2 \\
                    \end{cases}
                    \label{eqn:application__physical-state__reality}
                \end{equation}
            \end{minipage}
            & \begin{minipage}{3.8cm}
                \begin{equation}
                    \begin{aligned} & \land u_e' = u_\textrm{in} \\
                        & \land \lor \, s' = s \\ 
                        & \quad \lor s' = s - \delta \\
                    \end{aligned}
                    \label{eqn:application__physical-state__abstraction}
                \end{equation}
            \end{minipage}
    \end{tabular}
\end{table}

\noindent \textbf{Observational data ($O$)} The observational data, denoted as $\bm{O} = O_1 \ldots O_M$, are noisy, timestamped sensor measurements of the UAV's structural health, taken by $M$ sensors, indexed as $m = 1 \ldots M$. Sensor measurements inherently vary slightly from the actual structural health and each other due to sensor precision and environmental interference. Empirical data (sample shown in \eqref{eqn:application__observational-data__reality}), show typical small deviations from the ground truth value $S$.

In our abstraction, each sensor measurement $O_m$ is represented as the UAV's structural health value perturbed by some nondeterministic noise $\epsilon \in \{-1, 0, 1\}$.  \\

\begin{table}[h]
  \caption{Transition $S \rightedge O_m$: Generate observational data }
  \label{tab:application__observational-data}
  \centering
    \begin{tabular}{P{3.9cm}|P{3.8cm}}
        \textbf{Real-world process} & \textbf{Abstraction} \\
        \midrule
            \begin{minipage}{3.9cm}
                \begin{equation}
                    \begin{aligned}
                    o_m - h = [ & 0.57, 0.66, \\
                    & 0.31, -0.12, \\
                    & 0.42, -0.93, \\
                    & 0.61, \ldots ]
                    \end{aligned}
                    \label{eqn:application__observational-data__reality}
                \end{equation}
            \end{minipage}
            & \begin{minipage}{3.8cm}
                \begin{equation}
                    \begin{aligned} & o_m' = s + \epsilon \\
                    \end{aligned}
                    \label{eqn:application__observational-data__abstraction}
                \end{equation}
            \end{minipage} 
    \end{tabular}
\end{table}

\noindent \textbf{Distributed messaging of observational data} ($N$, $O_{m, \textrm{in}}$) In the UAV \dt{}, the communication of observational data via Bluetooth introduces complexities due to the potential unreliability of the wireless channels. 

In our abstraction, we construct separate variables for each Bluetooth channel ($N_1 \ldots N_M$) and for the received data $ O_{1, \textrm{in}} \ldots O_{M, \textrm{in}}$.

\begin{table}[h]
  \caption{Transition $O_m \rightedge N_m$: Transmit observational data}
  \label{tab:application__transmit-observational-data}
  \centering
    \begin{tabular}{P{3.1cm}|P{4cm}}
        \textbf{Real-world process} & \textbf{Abstraction} \\
        \midrule
            \begin{minipage}{3.1cm}
                \hspace{.5cm}
                \includegraphics[width=2.5cm]{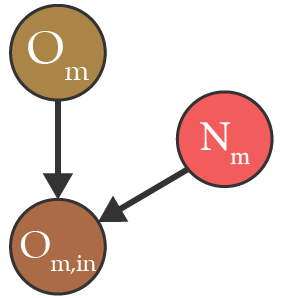}
            \end{minipage}
            & \begin{minipage}{4cm}
                \begin{equation}
                    n_m' =  \texttt{push} \; n_m, o_m 
                    \label{eqn:application__transmit-observational-data__abstraction}
                \end{equation}
            \end{minipage}
    \end{tabular}
\end{table}

Because our concern is at a higher level than the details of sensor and transmission operations, we treat the processes of data generation, abstracted in \eqref{eqn:application__observational-data__abstraction}, and transmission, abstracted in \eqref{eqn:application__transmit-observational-data__abstraction}, as mutually-atomic. This abstracts the generation and immediate transmission of data as a single, indivisible operation, as in \eqref{eqn:application__transmission-observational-data-combined}:

\begin{equation}
    \begin{split}
        \texttt{ObserveEmitObsAtomic} \coloneq & \; (O_m \rightedge N_m) 
         \land (S \rightedge O)
    \end{split}
    \label{eqn:application__transmission-observational-data-combined}
\end{equation}
The received value of a sensor message $O_m$ is represented by variable $O_{m, \textrm{in}}$. Per our methodology in Appendix~\ref{appendix:dt-state-machine__abstraction}, to model the unreliable receiving of messages, we remove the element at randomly-chosen index $i$ from queue $n_m$, and we add it to the received messages collection $o_{m, \textrm{in}}$. We impose the strong fairness condition that the correct message ($i=1$) is always eventually delivered. 

\begin{table}[h!]
  \caption{Transition $(O_m, N_m) \rightedge O_{m, \textrm{in}}$: Receive observational data}
  \label{tab:application__receive-observational-data}
  \centering
    \begin{tabular}{P{3.1cm}|P{4.5cm}}
        \textbf{Real-world process} & \textbf{Abstraction} \\
        \midrule
            \begin{minipage}{3.1cm}
              \hspace{.5cm}
              \includegraphics[width=2.5cm]{assets/application__transmit-observational-data.png}
            \end{minipage}
            & $\begin{aligned}  & \textrm{Let} \; i \in [1, \eta] \\
                & \land \texttt{Remove} \; n_m[i] \\
                & \land o_{m, \textrm{in}}' = \texttt{push} \; o_{m, \textrm{in}}, n_m[i] \\ 
                & \land SF(i=1) \\
                & \textrm{where} \; 1 \leq \eta \leq \textrm{Length} \; n_m
            \label{eqn:application__receive-observational-data}    
            \end{aligned}$ \\
    \end{tabular}
\end{table}

\noindent \textbf{Digital state} ($D$) The digital state $D$ represents the estimated structural health of the UAV, modeled as a variable within the range $\{1 \ldots 100\}$. This estimation is computed by a black-box model $\psi$, which outputs a predictive distribution for $D$. While the internal computations of each model remain undisclosed, output characteristics are discovered through prior statistical analysis.

Our abstraction retains the dependency of $D$ on previous state $D^{t-1}$, last control computed $U^{t-1}$ and the latest observational data $\bm{O^{t}_\textrm{in}}$. To enhance the model's tractability, we use known characteristics of $\phi$ to constrain the number of possible states for $D$. For instance, when analyzing the conditional probability for $D_O$, shown in \eqref{eqn:application__digital-state__reality},  where $D$ varies with $\bm{O^{t}_\textrm{in}}$ while keeping other factors constant, we observe that non-positive sensor observations significantly widen the range of possible values for $D$. Otherwise, $D$ typically fluctuates within a normal distribution $\mathcal{N}(d, \sigma^2)$, where the variance $\sigma^2$ is influenced by the type of control executed. To keep the abstraction tractable and focused on the most critical scenarios, we constrain $D$ to fluctuate within two standard deviations of the mean. This constraint is reflected in our abstraction, where $\zeta_2 \in [-1, 1]$ and $\zeta_3 \in [-5, 5]$ are set to represent the two standard deviation bounds for controls $u=2$ and $u=3$, respectively, rounded to the nearest integers.

\begin{table}[h!]
  \caption{Transition $(D, U, \bm{O_{m,\textrm{in}}}) \rightedge D$: Update digital state}
  \label{tab:application__digital-state}
  \centering
    \begin{tabular}{P{4.7cm}|P{3.1cm}}
        \textbf{Real-world process} & \textbf{Abstraction} \\
        \midrule
            \begin{minipage}{4.7cm}
                \begin{equation}
                    \begin{aligned}
                        &D \sim \psi(D^{t-1}, U^{t-1}, \bm{O^t_\textrm{in}}) \\
                        &D_O \sim \begin{cases}
                            \textrm{U}(0, d) & \exists o_i \leq 0 \\
                            \mathcal{N}(d, \sigma^2) & \textrm{otherwise}
                        \end{cases}
                    \end{aligned}
                    \label{eqn:application__digital-state__reality}
                \end{equation}
            \end{minipage}
    
            & 
            \begin{minipage}{3.5cm}
                \begin{equation}
                    \begin{aligned}  & \texttt{IF} \; \exists \, o_{m,\textrm{in}} : o_{m,\textrm{in}} \leq 0 \\
                    & \quad  d' = 0 \ldots d \\
                    & \texttt{ELSE} \\
                    & \quad \texttt{IF} \; u = 2 \\
                    & \qquad  d' = d + \zeta_2  \\
                    & \quad \texttt{ELSE} \\
                    & \qquad  d' = d + \zeta_3 \\
                    \end{aligned}
                    \label{eqn:application__digital-state__abstraction}
                \end{equation}
            \end{minipage}
    \end{tabular}
\end{table}

\noindent \textbf{Control ($U$)} The control $U$ is a command that instructs the UAV to execute either a 3\textit{g} or 2\textit{g} turn. The control is computed via a optimization model. 

In our abstraction (Table~\ref{tab:application__control}), the control $U$ is simplified to decision-making criteria based primarily on the UAV's estimated structural health $D$. This simplification is grounded in a prior analysis of the optimization model's outputs \cite{kapetyn2021}, which reveal that the value of 
$D$ primarily dictates whether the control $U$ can be set to 3.

\begin{table}[h!]
  \caption{Transition $D \rightedge U$: Compute and transmit control}
  \label{tab:application__control}
  \centering
    \begin{tabular}{P{4cm}|P{3.5cm}}
        \textbf{Real-world process} & \textbf{Abstraction} \\
        \midrule
            \begin{minipage}{4cm}
                \begin{equation}
                    u \in \begin{cases}
                        \{3\} & d \geq D_\textrm{min} \\
                        \{2,3\} & \textrm{otherwise}
                    \end{cases}
                \end{equation}
            \end{minipage}
            & $\begin{aligned}  & \texttt{IF} \; d \geq D_\textrm{min} \\
                & \enspace \lor u' = 3 \\
                & \enspace \lor u' = 2 \\
                & \texttt{ELSE} \\
                & \enspace  u' = 2 \\
            \end{aligned}$
    \end{tabular}
\end{table}

\noindent \textbf{Distributed messaging of control $(N_U, U_\textrm{in})$} Handling control messages is managed similarly to transmission and reception of observational data. In our abstraction (Table~\ref{tab:application__receive-control}), we assume that the correct message will eventually be delivered, and we treat the processes of computing a control decision and transmitting a control as mutually atomic operations, combined into a single, indivisible process to reduce complexity.

\begin{table}[h!]
  \caption{Transition $(U, N_u) \rightedge U_\textrm{in}$: Receive control}
  \label{tab:application__receive-control}
  \centering
    \begin{tabular}{P{3cm}|P{4.5cm}}
        \textbf{Real-world process} & \textbf{Abstraction} \\
        \midrule
            \begin{minipage}{3cm}
                \hspace{.5cm}
                \includegraphics[width=2.5cm]{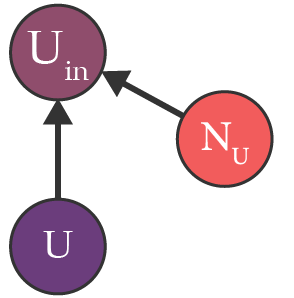}
            \end{minipage}
            & $\begin{aligned}  & \textrm{Let} \; i \in [1, \eta] \\
                & \land \texttt{Remove} \; n_u[i] \\
                & \land \texttt{IF} \; n_u[i].t > u_\textrm{in}.t \\
                & \qquad u_\textrm{in}' =  n_u[i] \\
                & \; \quad \texttt{ELSE} \\
                & \qquad u_\textrm{in}' =  u_\textrm{in} \\
                & \land SF(i=1) \\
                & \textrm{where} \; 1 \leq \eta \leq \textrm{Length} \, n_u
            \label{eqn:application__receive-control}    
            \end{aligned}$
    \end{tabular}
\end{table}

\subsubsection*{Termination}

Our UAV example uses specific termination conditions to reflect real mission parameters. Termination occurs when: (1) UAV reaches the maximum number of executed maneuvers $C_\textrm{max}$; (2) \dt{} exceeds a predefined maximum runtime $T_\textrm{max}$; or (3) \dt{} estimates the UAV’s structural health as non-positive, and all sensor readings concurrently indicate non-positive values, suggesting critical system failure.

\subsection{Specifying Properties}

\begin{figure}[h]
    \centering
    \includegraphics[width=3.5in]{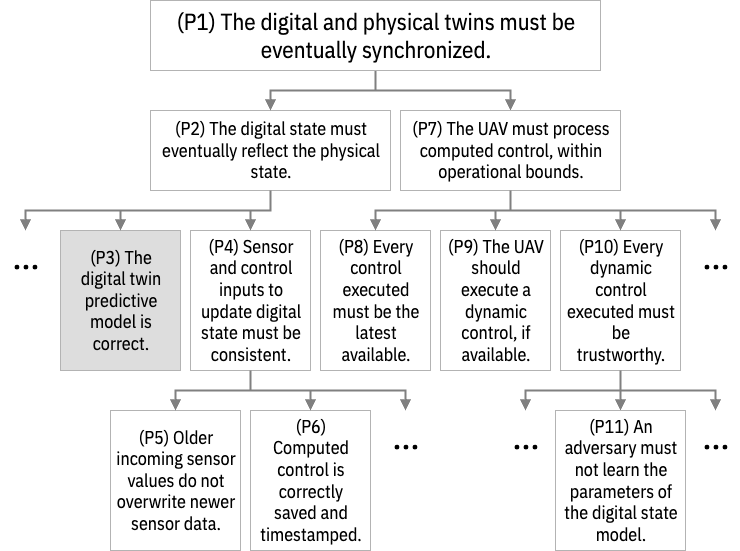}
    \caption{Partial property-part diagram showing a subset of properties}
    \label{fig:application__property-part-diagram}
\end{figure}

The core property of interest in the UAV \dt{} is \textit{synchronization}—the continuous, bidirectional feedback loop ensuring that the physical and digital entities reflect each other accurately. We define our primary synchronization property as:
\begin{center}
\textit{$P_1$: The physical and digital twins must be eventually synchronized.}
\end{center}
We use the term ``eventually'' to describe that synchronization will always be achieved, without binding it to a specific timeframe. To detail what synchronization entails, we deconstruct this overarching property into more granular sub-properties, guided by methodological questioning ---\textit{how}, \textit{what}, and \textit{why} \cite{dardenne1991, dardenne1993} --- with engineers and stakeholders. We discuss in more detail how we specify properties in Appendix~\ref{appendix__properties}.
The resulting property-part diagram, depicted in Fig.~\ref{fig:application__property-part-diagram}, illustrates a subset of these properties.

\subsection{Weakening Formal Verification with Statistical Guarantees}
\label{subsec:weakening}

Synchronization correctness require certain security properties to be satisfied. For example, in Fig.~\ref{fig:application__property-part-diagram}, $P_{11}$ requires that an adversary cannot infer information about the digital state model, which is necessary for the trustworthiness of messages exchanged between the physical and digital twins. This property falls under a class of security guarantees known as non-interference, a standard approach for formalizing information flow within a system. A process $r_1$ is noninterfering with another process $r_2$ across system $\mathcal{M}$ if $r_1$'s input to $\mathcal{M}$ has no effect on $M$'s output to $r_2$ \cite{mantel}. Different variations of noninterference exist \cite{nelson}, including generalized non-interference (GNI) which extends noninterference to probabilistic systems by mandating that for every pair of traces $b$ and $b'$, there exists a third trace $b''$ such that $b''$ agrees with the low-security inputs and $b''$ agrees with the high-security outputs \cite{McCullough}. The practicality of noninterference is well-known to be problematic \cite{KopfB2007}, and as of state-of-the-art, obeying GNI is still an impractical constraint on \dt{} systems. Here, we introduce a novel weakening of GNI with respect to particular secret \dt{} parameters, where we allow some information leakage while still maintaining formal bounds on the amount of relevant information leaked.

Notably, we measure information leakage through a \emph{system identification} perspective \cite{Ljung1999} rather than a generic information-theoretic view \cite{KopfB2007}, considering what systems-theoretic understanding of the \dt{} is leaked rather than just the number of bits about it, which may or may not be relevant to  adversarial action.  This is different from \cite{Chan_ea2022} which looks at state estimation rather than system identification, and \cite{LiuTYZ2022}, which is also quite different.

For example, consider the content of the communication involving the current health of the physical counterpart and the next action it is going to take. The change in health depends on the action taken and some system randomness. More concretely, let $h(t)\in\mathbb{N}\cup\{0\}$ denote the health of the system at time $t$. The system can take $m$ possible actions, indexed by $\{1,2,\ldots,m\}$.
Let $a(t)\in\{1,2,\ldots,m\}$ denote the action the system takes at time $t$. We assume the change in health $h(t)-h(t+1)$ is a Poisson random number drawn with rate $\lambda_{a(t)}$,
independent of all other changes in health:
\begin{align*}
    h(t)-h(t+1)=-\Delta h(t+1) \sim \text{Poisson}(\lambda_{a(t)}).
\end{align*}
This essentially implies that the health model of the system is given by $(\lambda_1,\lambda_2,\ldots,\lambda_m)$.

An adversary intercepts the communication between the \dt{} and the physical counterpart, and knows the values of $h(t)$ and $a(t)$. We want to determine whether the system's health model is compromised by this information leakage. So the estimation problem here is that given $\{(h(\tau),a(\tau)):1\le\tau\le t\}$, we want to figure out $\lambda_1,\lambda_2,\ldots,\lambda_m$.

In the general case, let us assume that $\lambda_i$ and $\lambda_j$ have no relation to each other (this may not be very practical since we often know which actions are costlier than others, but let us nevertheless make this simplifying assumption). So for estimating each $\lambda_i$, we only consider the set of times $\{\tau:a(\tau)=i\}$. In the absence of any prior, a good estimator for this would be
\begin{align}
    \hat{\lambda}_i = -\frac{1}{|\{\tau:a(\tau)=i\}|}\sum_{t\in\{\tau:a(\tau)=i\}}\Delta h(t+1).
    \label{eq:estimator}
\end{align}
Using standard probability results, this estimator has the following properties.
\begin{theorem}
    The estimator in \eqref{eq:estimator} satisfies
    \begin{enumerate}
        \item[(i)] $\mathbb{E}\left[\hat{\lambda}_i\right]=\lambda_i$.
        \item[(ii)] $\mathbb{P}\left(\lvert\hat{\lambda}_i-\lambda_i\rvert\ge\epsilon\right)\le\frac{\lambda_i}{N_i\epsilon^2}$,
        where $N_i=|\{\tau:a(\tau)=i\}|$.
    \end{enumerate}
\end{theorem}
Thus as the number of times a particular action is taken increases, we get a more accurate
estimate of the hit to health from that action and so we directly get a statistical guarantee on information leakage about the system properties.
In general, finite-sample bounds from system identification theory \cite{VenkateshD2001, SarkarRD2021, JonesD2022} can  characterize such \dt-relevant information leakage. With this weakening approach, we are able to satisfy property $P_{11}$, which would otherwise fail with a purely model-checking approach.

\section{Evaluation}
\label{sec:evaluation}

Our baseline specification for the UAV \dt{} encompasses various parameters: $M=2$ sensors, each with a maximum message delay of $\eta = 2$, a total of $C_\textrm{max} = 3$ possible mission maneuvers, and a maximum system runtime of $T_\textrm{max} = 4$. This specification manifests as 15 distinct processes and 18 variables, including auxiliary variables for supporting property verification, with 25 properties covering core system behavior. The TLA code closely mirrors the abstraction models presented in Sec.~\ref{sec:application}. An example code listing is shown in Appendix~\ref{appendix:code-listing}.

\subsection{Model Checking the State Space}

\begin{figure}[t]
    \centering
    \includegraphics[width=4in]{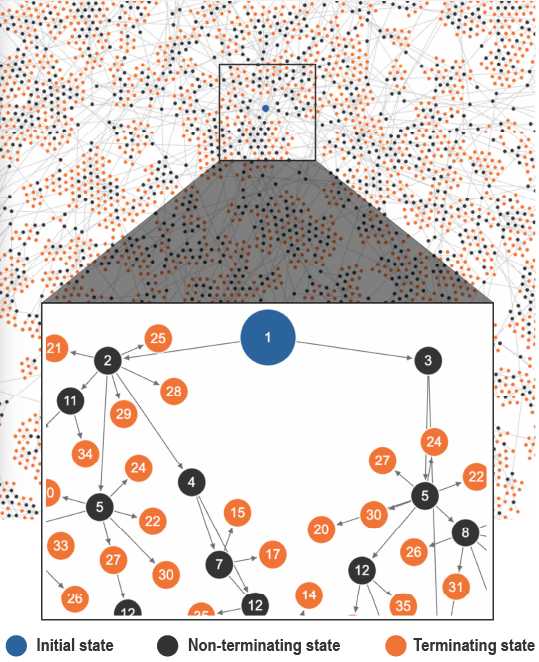}
    \caption{Visualization of state space: nodes represents states and edges represent transitions from state to state}
    \label{fig:evaluation__state-space}
\end{figure}

The state space generated by the UAV \dt's specification is combinatorially large, as each distinct process introduces a different potential interleaving, with every variable within these interleavings capable of assuming various values. We visualize this state space as a directed acyclic graph (DAG) in Fig.~\ref{fig:evaluation__state-space}, where each vertex represents a unique state---specific values assigned to variables---and edges depict transitions between these states. This graph is inherently a DAG, as it includes a model-checked guarantee of termination. In Fig.~\ref{fig:evaluation__state-space}, terminating states are highlighted in orange, while ongoing states are in black. The graph's initial state, depicted as a blue vertex (\texttt{1}), bifurcates into two principal pathways: the physical twin’s processes (\texttt{2}) and the \dt's processes (\texttt{3}). To highlight one possible pathway: from state (\texttt{2}), the system progresses to state (\texttt{4}) and then to (\texttt{7}), culminating in state (\texttt{15}). This final state indicates termination triggered by the UAV achieving the prescribed number of maneuvers.

Model checking is resource-intensive due to the vast size of the state space. On a hardware setup with 10 cores and 16 GB of RAM allocated to the TLC model checker, completing a single baseline model checking session requires approximately 15 hours. To evaluate scalability, we vary model parameters individually while keeping others constant. Increasing the number of sensors or the permissible message delay notably expands the state space by introducing more potential message interleavings. For example, with two sensors and a maximum message delay of $\eta=2$, the model generates $\num{13 534 045}$ states. Expanding to three sensors increases the state space to $\num{41 966 573}$, requiring two days to check on our hardware. We also examine the impact of atomicity assumptions by modifying the process $(S,U) \rightarrow S$, which asserts that the execution of control and the incurring of damage occur atomically (see Table~\ref{tab:application__physical-state}). By splitting the process into two interleaved, non-atomic steps, we unexpectedly observe a significant reduction in the state space --- from $12$ million to just one million distinct states. We hypothesize that this decrease results from the model checker simplifying invariants and pruning redundant states more effectively. This finding indicates that atomic assumptions do not always lead to larger state spaces and, in some cases, may simplify specification design. Table~\ref{tab:evaluation__model-params-state-space} summarizes the impact of varying model parameters.

\begin{table}[t]
    \centering
    \caption{Model parameters impact state space complexity}
    \begin{tabular}{l|r|r}
       \textbf{Specification} & \raggedleft \textbf{Distinct States} & \textbf{Total States} \\
       \midrule
       \raggedright Baseline & \num{12551574} & \num{33 960 246} \\
        \hspace{.1cm} $+1$ health ($S = 3$) & \num{24 668 110} & \num{66 833 826} \\
       \hspace{.1cm} $+1$ sensor ($M=3$) & \num{13 534 045} & \num{41 966 573} \\
       \hspace{.1cm} $+1$ delay ($\eta = 3$) & \num{15 307 358} & \num{50 720 696} \\
       \hspace{.1cm} $\pm 1$ noise ($\epsilon = \pm2$) & \num{15 804 834} & \num{42 619 510} \\
       \hspace{.1cm} $+1$ process ($|\Omega|+1$) & \num{1 227 202} & \num{3 231 322} \\
    \end{tabular}
    \label{tab:evaluation__model-params-state-space}
\end{table}

\subsection{Safety and Liveness Violations}

Throughout the development of our specification, we used an iterative approach that, while refining the design, also continuously exposed gaps that led to property violations. For instance, during a model checking session, we encountered a violation of property $P_8$: \textit{The executed command must be the latest command seen thus far}, formalized as $\Box [u_\textrm{executed} \neq \varnothing \land u_\textrm{executed}' \neq \varnothing \implies u_\textrm{executed}'.t > u_\textrm{executed}.t]$. The sequence of state transitions leading to this violation, simplified for clarity, includes the following key steps:
\begin{enumerate}
    \item \textbf{Initial State}: The system begins in its initial configuration.
    \item \textbf{Execute Command}: The UAV executes a backup command (timestamped $t=1$) because no dynamic command is available. 
\begin{lstlisting}[language=dict]
u_executed = {t: 1, name: "Backup", type: 2 }
\end{lstlisting}

    \item \textbf{Compute and Emit Command}: The \dt{} computes and emits a dynamic command (timestamped $t=1$).
\begin{lstlisting}[language=dict]
u = {t: 1, name: "Dynamic", type: 3 }
n_u = [{t: 1, name: "Dynamic", type: 3}] 
\end{lstlisting}

    \item \textbf{Receive Command}: The UAV receives this latest computed dynamic command.
\begin{lstlisting}[language=dict]
u_in = {t: 1, name: "Dynamic", type: 3 } 
n_u = []
\end{lstlisting} 

    \item \textbf{Execute Command}: The UAV executes the received dynamic command, violating $P_8$, as the command (timestamped $t=1$) is stale and should not have been executed.
\begin{lstlisting}[language=dict]
u_executed = {t: 1, name: "Dynamic", type: 3 }
\end{lstlisting} 
\end{enumerate}

\begin{figure}[t]
    \centering
    \includegraphics[width=4in]{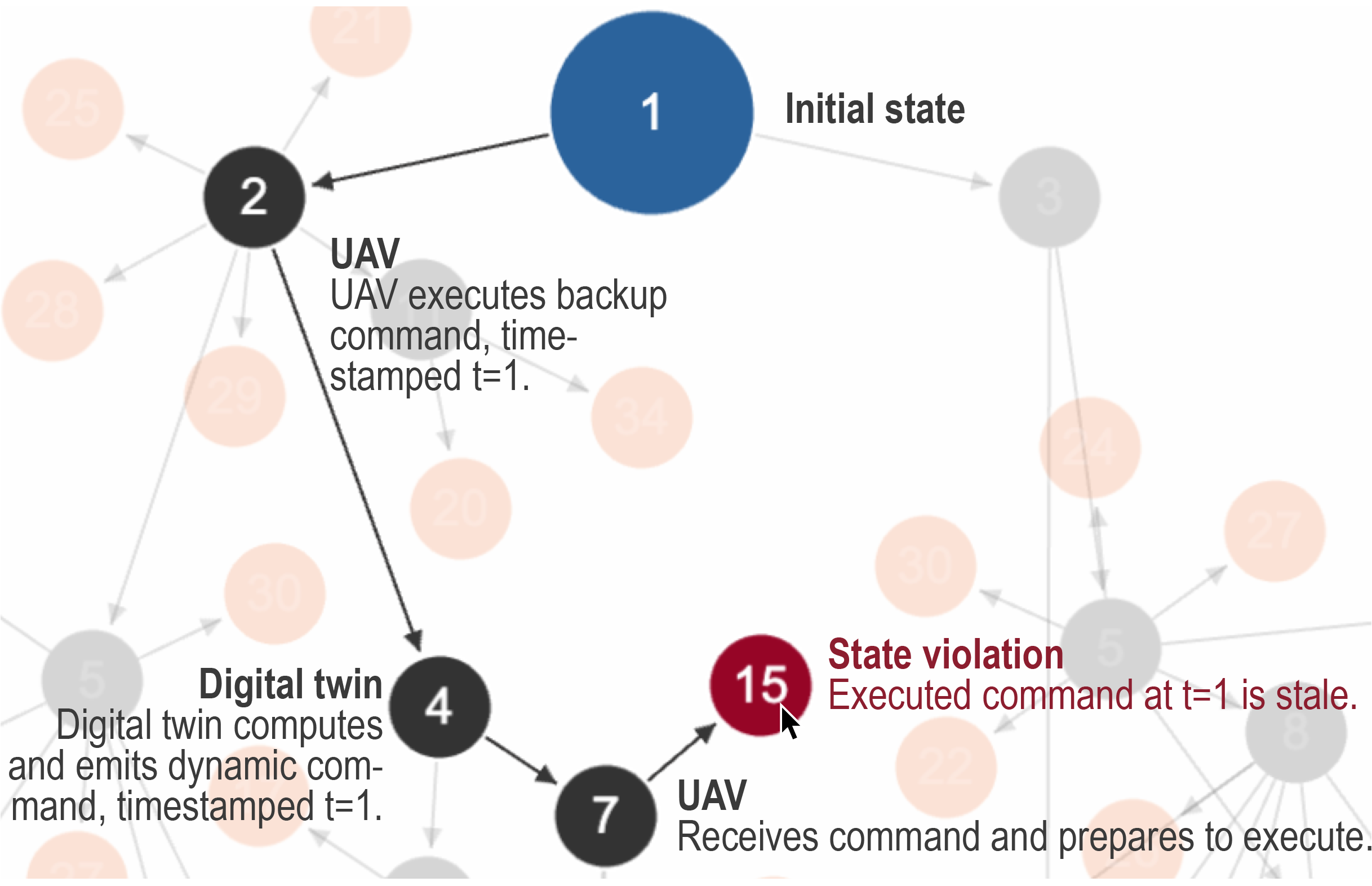}
    \caption{Graph visualization showing the path that leads to safety property violation}
    \label{fig:evaluation__property-violation}
\end{figure}

The progression of states leading to this violation is depicted in Fig.~\ref{fig:evaluation__property-violation}, where state (\texttt{15}) represents the state where the violation occurs. This issue stems from a critical oversight in the \textbf{Receive Command} process, where we had failed to implement a timestamp validation check for incoming command messages before their acceptance into the $u_\textrm{in}$ variable. While this oversight might seem straightforward to address in hindsight, it was easily overlooked during the initial stages of specification development. Fig.~\ref{fig:evaluation__before-after} shows the specification pre- and post-fix. This example underscores the importance of our iterative specification and model checking approach, particularly as design complexity increases, where seemingly fixes fixes can become obscured and go unnoticed.

\begin{figure}[t]
    \begin{subfigure}{1\linewidth}
    \begin{lstlisting}[mathescape=true]
PT_ReceiveControlDelayed(m_idx) ==
    /\ IF (s > 0 /\ u_executed_count <= MaxManeuvers)
       THEN /\ u_in' = n_u[m_idx]
            /\ n_u' = Remove(n_u, m_idx) 
    \end{lstlisting}
    \caption{Previously: Transition action for UAV receiving commands violates property $P_8$}
    \end{subfigure}
    \begin{subfigure}{1\linewidth}
    \begin{lstlisting}[mathescape=true]
PT_ReceiveControlDelayed(m_idx) ==
    /\ IF (s > 0 /\ u_executed_count <= MaxManeuvers)
       THEN /\ $\texttt{\color[HTML]{0000ff}{IF n\_u[m\_idx]["t"] > u\_in["t"]}}$
               THEN /\ u_in' = n_u[m_idx]
                    /\ n_u' = Remove(n_u, m_idx)
    \end{lstlisting}
    \caption{After: Transition action for UAV receiving commands with addition of a timestamp check}
    \end{subfigure}
    \caption{Example of a property violation and subsequent fix}
    \label{fig:evaluation__before-after}
\end{figure}

\section{Discussion}
\label{sec:discussion}

This paper presents a methodology for developing formally verifiable DT designs using TLA by transforming the PGM framework into a finite state machine with an augmentation for distributed communication. This approach enables the abstraction of complex distributed DT dynamics, allowing for the verification of synchronization properties. Because traditional formal methods have limitations, particularly with strict security definitions, we address this with a novel weakening method that combines formal verification with statistical guarantees. This allows controlled information leakage while ensuring these weakened properties align with the property-part diagram used in model checking. Despite the challenge of state space explosion, a common issue in model checking \cite{clarke}, even models with small parameters revealed early design errors. This iterative process highlights the value of formal verification in safety-critical systems, and future work will focus on bridging the gap between high-level formal specifications and practical \dt{} implementations.

\section*{Acknowledgment}
We thank James Bornholt and Akhil Bhimaraju for their valuable insights and feedback.

\bibliographystyle{IEEEtran}
\bibliography{main}

\appendix

\section{Background on the PGM framework and TLA}
\label{appendix:background}

A PGM encodes random variables as nodes and statistical dependencies as edges between nodes. In the DT PGM framework, the PGM governs how variables are updated in each timestep. An edge between two nodes dictates the dependence of the destination node on the parent node. 
For example, the edge $S \rightedge O$ represents that observational data $O$ depends on the physical state $S$. 

\begin{figure}[h]
\centering
    \includegraphics[width=4.4in]{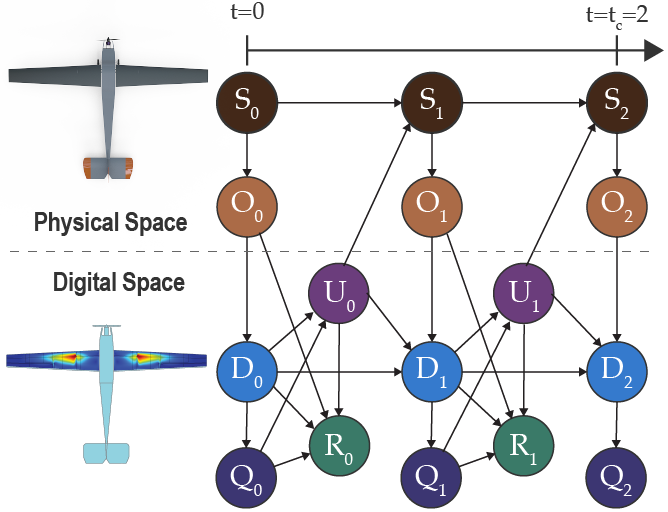}
    \caption{Probabilistic graphical model (PGM) describing the UAV DT.}
    \label{fig:appendix__pgm}       
\end{figure}

\section{State Machine Formalization}
\label{appendix:dt-state-machine}

This section details additional assumptions of the finite state machine formalization.

\subsubsection*{Fairness.}
\label{appendix:dt-state-machine__fairness}

Fairness is crucial in scheduling processes within the state machine framework. It ensures all processes are treated justly under the operational rules of the system. Specifically:
\begin{equation}
    \begin{split}
        \mathcal{F} = \{ \textit{UF}(\omega_i) \lor \textit{WF}(\omega_i) \lor \textit{SF}(\omega_i) \mid \omega_i \in \Omega \} 
    \end{split}
    \label{eqn:model__fairness}
\end{equation}
Here, \eqref{eqn:model__fairness} categorizes each process in the process set $\Omega$ into three types: Unfair (UF), where it is allowable that a process may never execute; Weakly Fair (WF), which guarantees that if a process is continuously enabled, it will eventually execute; and Strongly Fair (SF), requiring that if a process is enabled intermittently, it must eventually be executed. The classification into these categories depends on the system's dynamic requirements and stakeholder inputs, guiding how processes are triggered during the state machine's operation. For instance, a process like $S \rightedge S$, representing the continuous update of the UAV's physical state, is deemed Strongly Fair because its execution is essential and inevitable, reflecting the continuous nature of physical state updates. Weak fairness is generally assumed and warranted for many real-world systems \cite{brook, glabbeek} as it is unrealistic that a process can wait indefinitely before executing.

\subsubsection*{Initial and Termination States}\label{appendix:dt-state-machine__initial-termination}

The initial state of a state machine is crucial as it sets the baseline from which all processes begin. For the DT, this is defined by a logical predicate that assigns a starting value to each variable in the system. This predicate ensures all components of the DT start in a well-defined state that is consistent with the expected initial conditions of its physical twin. For instance, if modeling a UAV DT, the initial state may specify the UAV's starting health, the initial digital state, etc.

While a DT can theoretically operate indefinitely, practical applications often require defining specific conditions under which the simulation or operation should cease. This is termed the termination state $\mathcal{T}$, which is also expressed as a logical predicate. Termination conditions can vary widely depending on the system's purpose but generally include achieving a goal, exhausting resources, or encountering a specific event that requires halting operations. For example, a UAV's digital twin might terminate when the UAV completes its mission objectives or when the digital twin finishes program execution for a specified duration.

\subsection{Abstraction of Digital Twin complexity}\label{appendix:dt-state-machine__abstraction}

In developing the state machine for the DT, our aim was not only to capture the dynamic interplay of components but also to abstract complex DT behaviors into manageable and verifiable forms. This abstraction focuses on simplifying intricate component behaviors---whether physical phenomena or computational complexities---while preserving the essential characteristics necessary for accurate system modeling. Here, we outline our methodology for abstracting these complexities, providing general principles that are later applied in specific contexts within Sec.~\ref{sec:application}. \\

\noindent \textbf{Asserting existence instead of computing numerical operations} One fundamental principle in formal methods is to describe the outcomes of system operations rather than detailing the specific computations that achieve these outcomes \cite{lamport2002, luckcuck, rungger}. In line with this principle and consistent with prior literature in other domains \cite{Misra2011, Stanley-Marbell2021}, our approach simplifies the numerical complexity inherent in DT operations, yet maintains the integrity of the computational results in the abstracted state machine actions. \\

\noindent \textbf{Retain probabilistic characteristics} In our methodology, simplification does not eliminate  stochastic behavior inherent to many DT processes, especially those involving predictive, probabilistic components. However, instead of quantifying specific probabilities, our abstraction focuses on delineating possible behaviors by writing indeterministic actions in TLA \cite{lamport2002}. \\

\noindent \textbf{Representing a distributed message channel as a queue with deterministic write and nondeterministic read} Recall from Sec.~\ref{sec:dt-state-machine} that we represent a messaging channel as a random variable in the augmented PGM. Following the established practice in formal methods of using queues to represent network communication \cite{lynch}, we abstract each message channel random variable as a queue, where messages are `pushed' onto the queue as they are sent and `popped' from the queue in a nondeterministic order. This reflects potential real-world communication issues like delays, losses, or reordering, simplifying the analysis by eliminating the need to track details such as the timing specifics of each message.

\begin{equation}
    \begin{split}
        \textrm{channel} &\gets \textrm{Queue}[] \\
        \texttt{push}\; x &\coloneq \textrm{append channel}, x \\
        \texttt{pop i} &\coloneq \textrm{channel}[i] \\
        \texttt{remove i} &\coloneq \{\textrm{channel}[i] \in 1..\textrm{Length(channel)} \mid i \neq j  \} \\
        &  \textrm{where} \; i \in [1, \eta] \\
        & \textrm{with} \; SF(i=1) \\
    \end{split}
    \label{eqn:model__message-channel}
\end{equation}

\noindent Index $i$ ranges from $1$ to $\eta$, a parameter representing out-of-order message delivery constraints. For instance, a realistic and common constraint in wireless networks is a limited buffering and processing window. Index $i$ is randomly chosen, where $i=1$ indicates that the correct, most recent message is being delivered. We impose the strong fairness condition $SF(i = 1)$ to ensure the correct message will eventually be delivered, a realistic expectation that mirrors  guarantees provided by many network protocols such as Bluetooth. \\

\noindent \textbf{Atomicity in system orchestration} Atomicity is a fundamental consideration in formal methods, particularly when defining specifications for complex systems \cite{lamport1985}. In our context, two processes are deemed mutually atomic when their state transitions are considered simultaneous and uninterruptible, essentially occurring within the same atomic slice of time. In designing the orchestration for the DT, we strategically treat certain groups of operations as atomic to simplify the model and enhance tractability. This decision helps manage complexity by reducing the granularity of interaction between components, focusing on high-level system behavior rather than the minutiae of inter-process communication.

\subsection{Specifying Properties}
\label{appendix__properties}

In the process of specifying properties, initial analysis identifies two critical aspects of synchronization: (1) Consistency between the digital and physical states, and (2) Execution of control commands by the physical twin as issued by the DT. These insights lead to the development of properties $P_2$ and $P_7$, visualized in Fig.~\ref{fig:application__property-part-diagram}. Further decomposition of $P_2$ reveals that the digital state's update relies on consistent sensor data and computed control inputs, reflected in property $P_4$: \textit{Sensor and control inputs to update digital state must be consistent}. We observe that for consistent sensor inputs to be feasible, the DT must effectively receive and process incoming sensor data, a requirement captured in property $P_5$. This analysis continues, breaking down each goal into finer-grained sub-properties, eventually organizing them using goal structuring notation \cite{jackson}. 

Correct orchestration naturally requires component correctness as well. For example, property $P_3$ articulates that the predictive model within the digital twin must be correct to ensure that the digital state accurately twins the physical state over time. We describe these critical functionalities as explicit sub-properties within our framework. For instance, we configure our design to simulate scenarios where $P_3$ is false, allowing us to explore and verify the system's behavior under conditions of component failure. This transformation of potentially implicit assumptions into concrete, testable elements within our specification not only crystallizes the verifiable state of each component but also elucidates their role and relationship to the system's overall behavior.

\section{TLA Code}\label{appendix:code-listing}

The TLA code is written and model-checked with the TLA+ toolkit. Fig.~\ref{fig:appendix__code-snippet} shows a portion of this code, specifically modeling the procedure \texttt{DT\_ReceiveObsDelayed(s, m)}. This process represents actions taken for a specific sensor \texttt{s} with a particular delay index \texttt{m}.

In this snippet, \texttt{DT\_ReceiveObsDelayed} checks if there are any queued observations for sensor \texttt{s} (by verifying \texttt{n\_obs[s]} is not empty). If observations exist, it then checks if the delay index \texttt{m} corresponds to an entry within the domain of \texttt{n\_obs[s]}. If so, it evaluates whether the timestamp \texttt{n\_obs[s][m]["t"]} is recent enough, as determined by the function \texttt{OM!IsMessageUpToDate}. If the message is up-to-date, it appends this observation to the list of received observations \texttt{obs\_in[s]} and removes it from the observation queue \texttt{n\_obs[s]}.
If the timestamp is outdated, the entry is removed from \texttt{n\_obs[s]}, leaving \texttt{obs\_in[s]} unchanged. In cases where the delay index \texttt{m} is not present, both \texttt{obs\_in[s]} and \texttt{n\_obs[s]} remain unchanged.

The full TLA+ specification, including this process and others within the DT model, is published on https://github.com/luwen-huang/uav\_dt.

\begin{figure}
\begin{lstlisting}[mathescape=true]
DT_ReceiveObsDelayed(s, m) == 
    ...
    /\  IF n_obs[s] # << >>
        THEN
            IF m_idx \in DOMAIN n_obs[s]
            THEN
                IF OM!IsMessageUpToDate(obs_in[s], n_obs[s][m]["t"])
                THEN
                    /\ obs_in' = [obs_in EXCEPT ![s] = Append(obs_in[s], n_obs[s][m])]
                    /\ n_obs' = [n_obs EXCEPT ![s] = Network!RemoveElement(n_obs[s], m)] 
                ELSE 
                    /\ n_obs' = [n_obs EXCEPT ![s] = Network!RemoveElement(n_obs[s], m)] 
                    /\ UNCHANGED << obs_in >> 
            ELSE
                UNCHANGED << obs_in, n_obs >>
        ELSE
            UNCHANGED << obs_in, n_obs >>
    ...
    /\ UNCHANGED << z, z_obs_inputs, z_c_input, z_counter, c, c_counter, c_obs_inputs, pt_vars, n_c >> 
\end{lstlisting}
\caption{TLA code listing showing the process of receiving asynchronous sensor observations.}
\label{fig:appendix__code-snippet}
\end{figure}

\end{document}